\documentclass[conference]{IEEEtran}
\IEEEoverridecommandlockouts
\usepackage{cite}
\usepackage{amsmath,amssymb,amsfonts}
\usepackage{algorithmic}
\usepackage{graphicx}
\usepackage{textcomp}
\usepackage{multirow}
\usepackage{booktabs}
\usepackage{hyperref}
\usepackage{xcolor}
\def\BibTeX{{\rm B\kern-.05em{\sc i\kern-.025em b}\kern-.08em
    T\kern-.1667em\lower.7ex\hbox{E}\kern-.125emX}}
\begin{document}

\newcommand{\red}[1]{\textcolor{red}{#1}}

\title{An Explicit Consistency-Preserving Loss Function for Phase Reconstruction and Speech Enhancement}

\author{
    \IEEEauthorblockN{
        Pin-Jui Ku\IEEEauthorrefmark{1},
        Chun-Wei Ho\IEEEauthorrefmark{1},
        Hao Yen\IEEEauthorrefmark{1},
        Sabato Marco Siniscalchi\IEEEauthorrefmark{1}\IEEEauthorrefmark{2} and
        Chin-Hui Lee\IEEEauthorrefmark{1}
    }
    \IEEEauthorblockA{
        \IEEEauthorrefmark{1}Georgia Institute of Technology, USA\\
    }
    \IEEEauthorblockA{
        \IEEEauthorrefmark{2}Università degli Studi di Palermo, Italy\\
    }
}

\maketitle

\begin{abstract}
In this work, we propose a novel consistency-preserving loss function for recovering the phase information in the context of phase reconstruction (PR) and speech enhancement (SE). Different from conventional techniques that directly estimate the phase using a deep model, our idea is to exploit ad-hoc constraints to directly generate a consistent pair of magnitude and phase. Specifically, the proposed loss forces a set of complex numbers to be a consistent short-time Fourier transform (STFT) representation, i.e., to be the spectrogram of a real signal. Our approach thus avoids the difficulty of estimating the original phase, which is highly unstructured and sensitive to time shift. The influence of our proposed loss is first assessed on a PR task, experimentally demonstrating that our approach is viable. Next, we show its effectiveness on an SE task, using both the VB-DMD and WSJ0-CHiME3 data sets. On VB-DMD, our approach is competitive with conventional solutions. On the challenging WSJ0-CHiME3 set, the proposed framework compares favourably over those techniques that explicitly estimate the phase.

\end{abstract}
\section{Introduction}
\label{sec:intro}

Phase reconstruction (PR) of short-term Fourier transform coefficients (STFT) is an important challenge in speech enhancement (SE) \cite{Paliwal2011TheIO, Gerkmann2015Phase, Krawczyk2014STFT} and speech separation tasks \cite{Mowlaee2012Phase, Jonathan2013Consistent, du2016regression, Wang2019Deep}.
Traditional approaches have mainly focused on manipulating the noisy magnitude, leaving the noisy phase spectrogram unchanged \cite{Berouti1979, Lim1979Wiener, Boll1979, Paliwal1987, mcclellan2003signal, Xu2014, Xu2015, Siniscalchi2021}.
However, pairing estimated magnitidue spectrograms with noisy phase information can lead to magnitude distortion and introduce artifacts into the speech signal\cite{Paliwal2011TheIO, Gerkmann2012Mmse, Paliwal2005Usefulness, Mowlaee2013Phase}, which mainly arises from inconsistencies between magnitude and phase spectrograms.

The idea of the magnitude-phase leveraging consistency constraints during phase reconstructio dates back to the Griffin-Lim algorithm (GLA)~\cite{Griffin1984Signal}. GLA iteratively transforms the complex-valued spectrogram into a time-domain signal and converts it back to the short-time frequency domain. During each iteration, the magnitude information is preserved while the phase estimate is refined. Building upon GLA concepts, \cite{le2008explicit} derives an explicit consistency constraints for complex-valued spectrograms.
These constraints were used to develop a flexible phase reconstruction algorithm that was used to perform a real-time processing based on sliding-block analysis, avoiding the GLA iterative refinements.


The growing research on deep SE models that aim to estimate both the magnitude and phase of the clean spectrogram simultaneously highlights the significance of accurate phase estimation~\cite{fu2019metricGAN, yin2020phasen, yen2022cold, ai2023neural, lu2023mp, ku2023s4nd, chao2024investigation, zhang2024unrestricted}. To achieve phase estimation,  different phase losses have been proposed, for instance, complex-domain loss~\cite{yin2020phasen, ku2023s4nd}, cosine distance loss~\cite{takamichi2018phase}, time-domain loss~\cite{hu2020dccrn, yen2022cold}, and anti-wrapping loss~\cite{ai2023neural, lu2023mp}.
These losses assume the deep SE model directly estimates the phase. However, the original phase spectrogram is highly unstructured due to phase wrapping~\cite{Kulmer2015phase} and is extremely sensitive to time shifts~\cite{Masuyama2020phase}, making training with such losses particularly challenging. Another major drawback is that the SE model attempts to generate the original phase while ignoring other plausible phase solutions that could still produce perceptually good speech. For instance, two signals \(x(n)\) and \(-x(n)\) have the same magnitude spectrogram, but their phases differ by a constant \( \pi \). This is known as the sign indeterminacy problem, which highlights the existence of at least two viable phase solutions for a given magnitude. Zhang~\cite{zhang2024unrestricted} reached similar conclusions, demonstrating that phase shifts by an arbitrary constant do not degrade speech quality.

In this work, we show indeed that the quality of the generated speech signal can be preserved as long as magnitude and phase spectrogram are consistent. In order to generate consistent magnitude and phase pairs, we propose a novel loss function for phase spectrogram reconstruction, which is to be used during the SE deep model training phase.
Our SE deep model is no longer forced to explicitly estimate a single phase solution, overcoming the above-mentioned problems, such as sign indeterminacy, by focusing on the generation of phase spectrogram consistent with its magnitude.
While a loss function based on the consistency constraint is used in recent SE models\cite{lu2023mp, chao2024investigation} through consistency projection~\cite{Griffin1984Signal, wisdom2019differentiable}, a key difference is that these works still primarily rely on conventional phase loss functions, such as anti-wrapping or complex-domain loss, with the consistency-based loss treated merely as an auxiliary loss. In other words, the main objective remains the direct estimation of the original phase, leaving the challenges of phase estimation unresolved.
To the best of our knowledge, our paper is the first to introduce the consistency constraint as the only phase loss function to train SE models.
The feasibility of our approach is initially evaluated on the PR task by comparing it with existing methods that directly perform phase estimation. Experimental results validate the effectiveness of our technique. Next we consider the SE task,  using the VB-DMD and WSJ0-CHiME3 data sets.  We found that leveraging the proposed loss, the SE deep models attain consistent improvement in speech quality evaluation metrics.

\section{Related Techniques}
\label{sec:related_work}

Let \( \mathbf{H} \in \mathbb{C}^{M \times N} \) represent the complex-valued STFT spectrogram of the target signal with \( M \) time frames and \( N \) frequency bins. The magnitude spectrogram is denoted by \( \mathbf{A} = | \mathbf{H} | \), and the phase spectrogram by \( \mathbf{P} = \angle \mathbf{H} \). In the PR task, the deep model generates a phase estimate \( \mathbf{P}' \) given the magnitude \( \mathbf{A} \) as input. Different loss functions can be employed to achieve this task.

\noindent \textbf{Cosine Distance Loss:} It is based on the von Mises distribution~\cite{takamichi2018phase} and measures the distance between \( \mathbf{P} \) and \( \mathbf{P}' \) using the cosine distance~\cite{takamichi2018phase}:
\begin{equation}
    \mathcal{L}_{cos}(\mathbf{P}, \mathbf{P}') = - \sum_{m,n} \cos(\mathbf{P}_{m,n} - \mathbf{P}'_{m,n})
\end{equation}
Since the cosine function is periodic in \( 2\pi \), this loss function naturally addresses the phase wrapping issue.

\noindent \textbf{Complex-Domain Loss:} $L1$, or $L2$ distances have also been proposed to compute the distance between \( \mathbf{P} \) and \( \mathbf{P}' \) in the complex domain. For the $L2$ distance, the loss would be:
\begin{equation}
    \mathcal{L}_{comp\_L2}(\mathbf{P}, \mathbf{P}') = \sum_{m,n}\Vert \mathbf{H}_{m,n} - \mathbf{H}'_{m,n} \Vert_2^2
\end{equation}
where \( \mathbf{H}' = \mathbf{A}e^{j\mathbf{P}'} \). This loss can be interpreted as a cosine distance loss weighted by the magnitude \( \mathbf{A} \):
\begin{align}
     \sum_{m,n} \mathbf{A}_{m,n} \Vert e ^{j \mathbf{P}_{m,n}} &- e ^{j \mathbf{P}'_{m,n}}  \Vert_2^2 = \notag \\ 
     &\sum_{m,n} 2 \mathbf{A}_{m,n} (1 - \cos(\mathbf{P}_{m,n} -\mathbf{P}'_{m,n}))
\end{align}

\noindent \textbf{Time-Domain Loss:}  $L1$, or $L2$ distances can be computed in the time domain as well. The $L2$ loss in the time domain is:

\begin{equation}
    \mathcal{L}_{time\_L2}(\mathbf{P}, \mathbf{P}') = \sum_{n}\Vert \text{iSTFT} (\mathbf{H}) - \text{iSTFT} (\mathbf{H}') \Vert_2^2
\end{equation}
where the summation is over all time indices $n$ after applying the inverse STFT.

\noindent \textbf{Anti-wrapping Loss:} The anti-wrapping loss~\cite{ai2023neural} avoids  phase wrapping when computing the L2 distance between \( \mathbf{P} \) and \( \mathbf{P}' \) as follows:
\begin{align}
    \mathcal{L}_{AW}(\mathbf{P}, \mathbf{P}') =&\notag \\
    \sum_{m,n}\Vert \mathbf{P}_{m,n} - &\mathbf{P}'_{m,n}- 2\pi \cdot \text{round}\left(\frac{\mathbf{P}_{m,n} - \mathbf{P}'_{m,n}}{2\pi}\right) \Vert_2^2
\end{align}

\noindent \textbf{Loss Functions Based on Phase Derivatives:} It has been shown beneficial to define losses based on the distances between the target and predicted phase derivatives along the time and frequency axes, i.e., the group delay (GD), and the instantaneous frequency (IF)~\cite{ai2023neural, takamichi2018phase, zhang2024unrestricted}. Those quantities are more structured and less sensitive to time shifts compared to the phase itself. For example, the cosine distance loss can be modified as discussed in~\cite{takamichi2018phase,Masuyama2020phase}:
\begin{align}
    \mathcal{L}_{cos+derv}(\mathbf{P}, \mathbf{P}') & = \mathcal{L}_{cos}(\mathbf{P}, \mathbf{P}') \notag \\
    &  + \mathcal{L}_{cos}(\text{GD}(\mathbf{P}), \text{GD}(\mathbf{P}')) \notag \\
    & +  \mathcal{L}_{cos}(\text{IF}(\mathbf{P}), \text{IF}(\mathbf{P}'))
\end{align}
where \( \text{GD}(\cdot) \) and \( \text{IF}(\cdot) \) are functions that compute GD and IF from the phase, respectively. Similarly, the anti-wrapping loss can be modified as shown in ~\cite{ai2023neural}:
\begin{align}
    \mathcal{L}_{AW+derv}(\mathbf{P}, \mathbf{P}') & = \mathcal{L}_{AW}(\mathbf{P}, \mathbf{P}') \notag \\
    &  + \mathcal{L}_{AW}(\text{GD}(\mathbf{P}), \text{GD}(\mathbf{P}')) \notag \\
    & +  \mathcal{L}_{AW}(\text{IF}(\mathbf{P}), \text{IF}(\mathbf{P}'))
\end{align}

\section{Explicit Consistency-Preserving Loss}
\label{sec:method}
\subsection{Magnitude-Phase Consistency Constraints}

We define the following STFT quantities: window length \( N \), hop length \( R \), analysis window function \( W[n] \), and synthesis window function \( S[n] \).  \( W [n] \) and \( S[n] \) are zero outside the interval \( 0 \leq n \leq N - 1 \), and the window length \( N \) is an integer multiple of the hop length \( R \), with \( Q = \frac{N}{R} \).

For \( \mathbf{H} \) to be a consistent spectrogram, it must correspond to the STFT of a time-domain signal, which should also be, by definition,  the result of applying the inverse STFT to \( \mathbf{H} \). Thus, a necessary and sufficient condition for \( \mathbf{H} \) to be a consistent spectrogram is that it equals the STFT of its inverse STFT. Therefore, any consistent spectrogram \( \mathbf{H} \) must obey the following equation $\forall n' \in [0, N-1]$~\cite{le2008explicit}:
\begin{align}
\label{eq:ec_1}
\mathbf{H}_{m,n'} = &\frac{1}{N} \sum_k W[k] e^{-j2\pi k \frac{n'}{N}} \notag \\
( & S[k] \sum_{n=0}^{N-1} \mathbf{H}_{m,n} e^{j2\pi n \frac{k}{N}}  \notag \\
+ & \sum_{q=1}^{Q-1} S[k+qR] \sum_{n=0}^{N-1} \mathbf{H}_{m-q,n} e^{j2\pi n \frac{k+qR}{N}}  \notag \\
+ & \sum_{q=1}^{Q-1} S[k-qR] \sum_{n=0}^{N-1} \mathbf{H}_{m+q,n} e^{j2\pi n \frac{k-qR}{N}} )
\end{align}
By introducing the coefficients

\begin{equation}
    \alpha_q^{(R)}(p)=\frac{1}{N}\sum_k W[k]S[k+qR]e^{-j2\pi p\frac{k+qR}{N}}-\delta_p\delta_q
\end{equation}
where \( -(N-1) \le p \le N-1 \), \( -(Q-1) \le q \le Q-1 \), and \( \delta_i \) is the Kronecker delta, \cite{le2008explicit} simplifies Eq.~\ref{eq:ec_1} and shows that \( \mathbf{H} \in \mathbb{C}^{M \times N} \) is a consistent spectrogram if and only if

\begin{equation}
    \label{eq:ec_2}
    \sum_{q= -(Q-1)}^{Q-1} e^{j2\pi \frac{qR}{N} n}(\alpha_q^{(R)} * \mathbf{H})_{m-q, n} = 0
\end{equation}
is satisfied for all \( m \in [0, M-1] \) and for all \( n \in [0, N-1] \), where the convolution acts on the frequency axis of \( \mathbf{H} \).

\subsection{Proposed Explicit Consistency Loss}
Eq.~(\ref{eq:ec_2})
can be used to derive a loss function \( \mathcal{L}_{EC} \) that measures the consistency of \( \mathbf{H} \) as follows:

\begin{align}
    \mathcal{L}_{EC}(\mathbf{H}) = \sum_{m,n} \left| \sum_{q= -(Q-1)}^{Q-1} e^{j2\pi \frac{qR}{N} n}(\alpha_q^{(R)} * \mathbf{H})_{m-q, n} \right|^2
\end{align}
which is the reconstruction criterion employed in~\cite{le2008explicit}.
Thus, \( \mathcal{L}_{EC}(\mathbf{H}) \) can directly serve as the optimization criterion to train deep models generating complex-valued spectrograms. For instance, in the PR task, \( \mathcal{L}_{EC} \) can be used to force the phase prediction \( \mathbf{P}' \) such that \( \mathbf{H}' = \mathbf{A} e^{j \mathbf{P}'} \) is consistent:
\begin{align}
    \mathcal{L}_{EC}(\mathbf{P}') &= \notag \\
    \sum_{m,n} &\left| \sum_{q= -(Q-1)}^{Q-1} e^{j2\pi \frac{qR}{N} n}(\alpha_q^{(R)} * \mathbf{A}e^{j\mathbf{P}'})_{m-q, n} \right|^2
\end{align}
A key difference between \( \mathcal{L}_{EC}(\mathbf{P}') \) and the losses discussed in Sec.~\ref{sec:related_work} is that it does not involve any direct distance measurement between \( \mathbf{P} \) and \( \mathbf{P}' \). Therefore, the estimated phase spectrogram can deviate significantly from the original phase, as long as it results in a consistent spectrogram when combined with the magnitude spectrogram. \( \mathcal{L}_{EC}(\mathbf{P}') \) thereby implicitly handles the time shift issue and allows to explore a broader solution space that satisfies the consistency conditions. For instance, our loss function explains the observation by~\cite{zhang2024unrestricted} that a global phase shift does not degrade speech quality. If \( \mathbf{H} \) is a consistent spectrogram, then any spectrogram \( \mathbf{K} = \mathbf{H} e^{j\theta} \), with a global phase shift \( \theta \), remains consistent, as
\begin{gather}
    \label{eq:ec_3}
    \sum_{q= -(Q-1)}^{Q-1} e^{j 2 \pi \frac{qR}{N} n} (\alpha_q^{(R)} * \mathbf{K})_{m-q, n} = \notag \\
    e^{j\theta} \left( \sum_{q= -(Q-1)}^{Q-1} e^{j 2 \pi \frac{qR}{N} n} (\alpha_q^{(R)} * \mathbf{H})_{m-q, n} \right) = 0
\end{gather}
for all \( m, n \). Since the consistency of the spectrogram is unaffected by phase shifts, the speech quality remains intact.

\begin{figure}
    \centering
    \includegraphics[width=\linewidth]{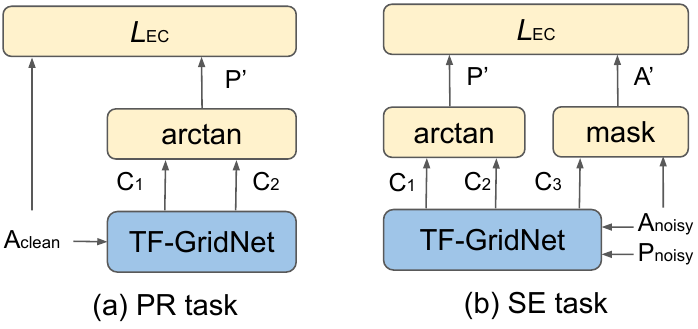}
    \caption{Illustration of inputs and outputs in our PR and SE experiments. Both $\mathbf{C}_1, \mathbf{C}_2 \in \mathbb{R}^{M \times N}$, and $\mathbf{C}_3 \in [0, 1]^{M \times N}$ are the model's outputs.}
    \vspace{-3mm}
    \label{fig:illustration}
\end{figure}

\section{Experimental Setup \& Results}

\subsection{Experimental Data Sets}
We use the \mbox{VoiceBank-DEMAND} (\mbox{VB-DMD})~\cite{ValentiniBotinhao2017} and \mbox{WSJ0-CHIME3}~\cite{lemercier2023storm} data sets in our experiments. For \mbox{VB-DMD}, we use the conventional train, validation, and test splits from \cite{Lu2022conditional}, resulting in 10,802 training, 770 validation, and 824 evaluation utterances. The WSJ0-CHiME3 data set was prepared as described in~\cite{lemercier2023storm}, with 12,776 training, 1,206 validation, and 651 test utterances. The \textit{Remix} and \textit{BandMask} data augmentation methods proposed in \cite{Dfossez2020real} are adopted. Our evaluation metrics include wide-band perceptual evaluation of speech quality (PESQ)~\cite{Rix2001}, extended short-term objective intelligibility (ESTOI)~\cite{Jensen2016stoi}, prediction of signal distortion (CSIG), and prediction of overall speech quality (COVL).

\subsection{Inputs and Outputs in the PR and SE Experiments}
TF-GridNet~\cite{Wang2023tfgridnet} is the backbone model used across all experiments. It was implemented with hyperparameters \( D=64 \), \( H=128 \), \( I=J=1 \), and \( B=3 \) as in~\cite{Wang2023tfgridnet}, excluding the global attention layer. This results in a model with approximately 1.3 million trainable parameters. 

Fig.~\ref{fig:illustration} illustrates the inputs and outputs adopted in TF-GridNet for the PR and SE experimental tasks. In the ``cheating'' PR task assuming the magnitude of clean speech \( \mathbf{A}_{clean} \)  is known, the model has two outputs, \( \mathbf{C}_1 \) and \( \mathbf{C}_2 \in \mathbb{R}^{M\times N} \), to compute the phase prediction \( \mathbf{P}' \) as
\(
\mathbf{P}' = \arctan \left( \frac{\mathbf{C}_1}{\mathbf{C}_2} \right)
\). The predicted phase \( \mathbf{P}' \) is either compared with the original phase of clean speech \( \mathbf{P}_{clean} \) using one of the losses in Sec.~\ref{sec:related_work}, or combined with the clean magnitude \( \mathbf{A}_{clean} \) as indicated in the proposed loss function \( \mathcal{L}_{EC} \). As in the ``real'' SE task in Fig.~\ref{fig:illustration}(b) using noisy magnitude and phase spectrograms \( (\mathbf{A}_{noisy}, \mathbf{P}_{noisy}) \) as inputs, the model produces three outputs, with the third \( \mathbf{C}_3 \) used as a real-valued mask to generate the estimated magnitude spectrogram \( \mathbf{A}' \). The $L1$ loss is  used to compute the error between \( \mathbf{A}' \) and clean magnitude \( \mathbf{A}_{clean} \). Different from what done for PR, \( \mathbf{P}' \) is paired with \( \mathbf{A}' \) instead of $\mathbf{A}_{clean}$  when computing \( \mathcal{L}_{EC} \) in Eq.~(\ref{eq:ec_2}). 

The training uses 32-second batches for 50 and 160 epochs in the PR and SE tasks, respectively. The Adam optimizer with an initial learning rate of \( 1 \times 10^{-3} \), which decays to \( 1 \times 10^{-5} \) via cosine annealing, is used. STFT is computed with a 512-sample window with a 128-sample hop size of a Hann window function. Magnitude compression is then applied per \cite{lemercier2023storm} with the compression parameters \( a = 0.5 \) and \( b = 1 \).

\begin{table}[t]
\caption{Comparison of Different Phase Loss Functions on Phase Reconstruction Task}
\label{table:phase_reconstruction}
\vspace{-1.0em}
\begin{center}
  \setlength\tabcolsep{6.0pt} 
  \resizebox{0.87\columnwidth}{!}{%
    \begin{tabular}{cccccc}
      \toprule
      Loss                 & PESQ & ESTOI & CSIG & COVL  \\
      \midrule
      Noisy Phase              & 3.95 & 0.99 & 4.99 & 4.72  \\
      \midrule
      $\mathcal{L}_{cos}$          & 3.06 & 0.88 & 4.73   & 3.94    \\
      $\mathcal{L}_{AW}$           & 1.31 & 0.64 & 3.30   & 2.25    \\
      $\mathcal{L}_{comp\_L1}$     & 3.55 & 0.93 & 4.94   & 4.34    \\
      $\mathcal{L}_{comp\_L2}$     & 3.60 & 0.91 & 4.97   & 4.39    \\
      $\mathcal{L}_{time\_L1}$     & 3.06 & 0.84 & 4.43   & 3.79    \\
      $\mathcal{L}_{time\_L2}$     & 2.87 & 0.93 & 4.42   & 3.68    \\
      \midrule
      $\mathcal{L}_{EC}$           & 4.15 & 0.98 & 4.99   & 4.85   \\
      \midrule \midrule
      $\mathcal{L}_{AW+derv}$     & 4.34 & 0.99 & 5.00 & 4.97   \\
      $\mathcal{L}_{cos+derv}$    & 4.29 & 0.99 & 5.00 & 4.95   \\
      \bottomrule
    \end{tabular}
  }
   \vspace{-1.0em}
\end{center}
\end{table}

\subsection{Phase Reconstruction Results}
\label{subsec:pr_results}
We first consider the PR task on the VB-DMD data set. As demonstrated in Table~\ref{table:phase_reconstruction}, when the loss is calculated directly on phase spectrograms (i.e., without taking phase derivatives into account), deep models trained using any of the losses discussed in Sec.~\ref{sec:related_work} fail to produce phase spectrograms that result in high-quality speech. On the average, the speech quality is worse than that attained using the noisy phase as indicated in the top row of Table~\ref{table:phase_reconstruction}. This highlights the difficulty of direct phase estimation due to the randomness and time-shift sensitivity of the target phase. In contrast, when the deep model is trained with \( \mathcal{L}_{EC} \), the generated phase to reconstruct a speech signal indeed outperforms the quality attained with the noisy phase, achieving an average PESQ of 4.15. This result supports our hypothesis that the quality of the generated speech signal can be maintained as long as the magnitude and phase are consistent in order to effectively guide a deep model in the ``cheating'' PR task.

From Table~\ref{table:phase_reconstruction}, we can also observe that leveraging phase derivatives improves the deep training stage using either the cosine distance or the anti-wrapping loss resulting in a good phase estimation, as implicitly indicated by the better reported results shown in the two bottom rows in Table~\ref{table:phase_reconstruction}.
\begin{table}[t]
\caption{Comparisons of speech enhancement performances on the VB-DMD data set}
\vspace{-2.0em}
\label{table:enh_vbdmd}
\begin{center}
  \setlength\tabcolsep{4.0pt} 
  \resizebox{\columnwidth}{!}{%
    \begin{tabular}{cccccc}
      \toprule
      Loss & MetricGAN & PESQ & ESTOI & CSIG & COVL   \\
      \midrule
      Noisy Speech   &  --    & 1.97  & 0.72  & 3.49  & 2.74  \\
      \midrule
      $\mathcal{L}_{cos}$       & N & 3.37  & 0.89    & 4.70  & 4.13   \\
      $\mathcal{L}_{AW}$   & N & 3.36  & 0.89    & 4.70  & 4.13   \\
      $\mathcal{L}_{cos+derv}$ & N & 3.36  & 0.89    & 4.70  & 4.12   \\
      $\mathcal{L}_{AW+derv}$ & N & 3.37  & 0.89    & 4.71  & 4.14   \\
      $\mathcal{L}_{EC}$     & N & \textbf{3.39}  & 0.89    & \textbf{4.71}  & \textbf{4.15}   \\
      \midrule
      $\mathcal{L}_{cos}$       & Y & 3.49  & 0.89    & 4.74  & 4.22   \\
      $\mathcal{L}_{AW}$   & Y & 3.50  & 0.89    & 4.75  & 4.23   \\
      $\mathcal{L}_{cos+derv}$ & Y & 3.49  & 0.89    & 4.75  & 4.23   \\
      $\mathcal{L}_{AW+derv}$ & Y & 3.51  & 0.89    & \textbf{4.77}  & \textbf{4.25}   \\
      $\mathcal{L}_{EC}$     & Y & \textbf{3.53}  & 0.89    & \textbf{4.77}  & 4.23   \\
      \bottomrule
    \end{tabular}
  }
  \vspace{-1.0em}
\end{center}
\end{table}

Those results might call into question the need for the proposed loss function. However, we found that the benefits of incorporating phase derivative loss are only evident when the ground-truth magnitude information is available, which is in line with~\cite{Masuyama2020phase, Thieling2021recurrent}. For example, the clean magnitude is not available in the ``real'' SE task, and that makes an accurate phase derivative estimation much more challenging due to the coupling between phase and magnitude~\cite{Shimauchi2017on, wang2021compensation}. 
Furthermore, in the SE task the model is provided with noisy phase, which serves as a good intial estimate of the clean phase as shown in the first row in Table~\ref{table:phase_reconstruction} and alleviates the difficulty of minimizing the direct phase loss without considering phase derivatives.

\subsection{Speech Enhancement Results}
\label{subsec:se_results}
We now assess the use of the proposed loss in the SE task. Both the cosine distance and the anti-wrapping losses are used for comparison, with or without  phase derivatives. Furthermore,  we analyze the effect of including the MetricGAN loss~\cite{fu2019metricGAN}, which has proven useful in the SE task.

Table~\ref{table:enh_vbdmd} presents the evaluation results on the VB-DMD data set. It can be observed that the cosine distance and anti-wrapping losses lead to similar SE results. As expected, no improvement is observed when including a loss based on phase derivatives. In contrast, leveraging \( \mathcal{L}_{EC} \), a marginal improvement in PESQ is obtained independently of the MetricGAN loss. Remarkably, our results using \( \mathcal{L}_{EC} \) as the only phase loss are comparable to state-of-the-art solutions, such as MP-SENet~\cite{lu2023mp} and SEMamba~\cite{chao2024investigation}, which are trained using multiple phase loss functions with carefully fine-tuned weights for each individual loss.

To further test the effectiveness of the proposed loss function in a more challenging scenario, we conduct the same experiment on the \mbox{WSJ0-Chime3} data set, where SNR values can be as low as -6 dB. As shown in Table~\ref{table:enh_wsj0}, similar conclusions to those from Table~\ref{table:enh_vbdmd} can be drawn. Furthermore and more importantly, we found that the proposed consistency-preserving loss function provides an even more remarkable improvement on this more challenging data set, with a boost of around 0.7 PESQ scores regardless of the use of the MetricGAN loss. This result confirms our assumption that the proposed function serves as an ideal phase loss in the SE task by considering the consistency between the estimated magnitude and phase and facilitates the deep model to explore a broader solution space instead of trying to directly estimate the original phase.

\begin{table}[t]
\caption{Comparisons of speech enhancement performances on the WSJ0-Chime3 data set}
\vspace{-2.0em}
\label{table:enh_wsj0}
\begin{center}
  \setlength\tabcolsep{4.0pt} 
  \resizebox{\columnwidth}{!}{%
    \begin{tabular}{cccccc}
      \toprule
      Loss & MetricGAN & PESQ & ESTOI & CSIG & COVL  \\
      \midrule
      Noisy Speech
      &  --    & 1.35  & 0.63  & 3.13  & 2.27  \\
      \midrule
      $\mathcal{L}_{cos}$       & N & 2.95  & 0.91    & 4.46 & 3.74   \\
      $\mathcal{L}_{AW}$   & N & 2.98  & 0.91    & 4.51 & 3.79   \\
      $\mathcal{L}_{cos+derv}$      & N & 2.97  & 0.91    & 4.52 & 3.79   \\
      $\mathcal{L}_{AW+derv}$ & N & 3.00  & 0.91    & 4.54 & 3.82   \\
      $\mathcal{L}_{EC}$     & N & \textbf{3.04}  & 0.91    & \textbf{4.56} & \textbf{3.85}   \\
      \midrule
      $\mathcal{L}_{cos}$       & Y & 3.15  & 0.91    & 4.61 & 3.93   \\
      $\mathcal{L}_{AW}$   & Y & 3.14  & 0.91    & 4.59 & 3.92   \\
      $\mathcal{L}_{cos+derv}$ & Y & 3.15  & 0.91    & 4.58 & 3.91   \\
      $\mathcal{L}_{AW+derv}$ & Y & 3.15  & 0.91    & 4.62 & 3.94   \\
      $\mathcal{L}_{EC}$       & Y & \textbf{3.21}  & 0.91    & \textbf{4.63} & \textbf{3.97}   \\
      \bottomrule
    \end{tabular}
  }
  \vspace{-1.0em}
\end{center}
\end{table}



%
\section{Conclusion}
In this paper, we propose a novel loss function for phase estimation based on the magnitude-phase consistency constraint. Unlike existing phase losses, the proposed approach does not enforce a single-phase solution path. Instead, it requires only that the model generates a phase spectrogram consistent with its corresponding magnitude without specifying any details of the original phase, thus allowing for a broader solution space. Experimental results confirm the effectiveness of the proposed loss function, demonstrating superior performance compared to existing cosine distance loss and anti-wrapping loss, particularly in low-SNR scenarios.

%
\cleardoublepage
%
\bibliographystyle{IEEEtran}
\bibliography{references/refs}

\end{document}